\documentclass[preprint,pre,aps,showpacs,amsmath]{revtex4}
\usepackage{graphicx}

\begin{document}

\title{Frequency-Domain Order Parameters for the Burst And Spike Synchronization Transitions of Bursting Neurons}
\author{Sang-Yoon Kim}
\email{sangyoonkim@dnue.ac.kr}
\author{Woochang Lim}
\email{woochanglim@dnue.ac.kr}
\affiliation{Computational Neuroscience Lab., Department of Science Education, Daegu National University of Education, Daegu 705-115, Korea}

\begin{abstract}
We are interested in characterization of synchronization transitions of bursting neurons in the frequency domain. Instantaneous population firing rate (IPFR) $R(t)$, which is directly obtained from the raster plot of neural spikes, is often used as a realistic collective quantity describing population activities in both the computational and the experimental neuroscience. For the case of spiking neurons, a realistic time-domain order parameter, based on $R(t)$, was introduced in our recent work to characterize the spike synchronization transition.
Unlike the case of spiking neurons, the IPFR $R(t)$ of bursting neurons exhibits population behaviors with both the slow bursting and the fast spiking timescales. For our aim, we decompose the IPFR $R(t)$ into the instantaneous population bursting rate $R_b(t)$ (describing the bursting behavior) and the instantaneous population spike rate $R_s(t)$ (describing the spiking behavior) via frequency filtering, and extend the realistic order parameter to the case of bursting neurons. Thus, we develop the frequency-domain bursting and spiking order parameters which are just the bursting and spiking ``coherence factors'' $\beta_b$ and $\beta_s$ of the bursting and spiking peaks in the power spectral densities of $R_b$ and $R_s$ (i.e., ``signal to noise'' ratio of the spectral peak height and its relative width). Through calculation of $\beta_b$ and $\beta_s$, we obtain the bursting and spiking thresholds beyond which the burst and spike synchronizations break up, respectively. Consequently, it is shown in explicit examples that the frequency-domain bursting and spiking order parameters may be usefully used for characterization of the bursting and the spiking transitions, respectively.
\end{abstract}


\maketitle

\section{Introduction}
\label{sec:INT}

Recently, much attention has been paid to brain rhythms, observed in electrical recordings of firing activity \cite{Buz}. These brain rhythms emerge via synchronization between firings of individual neurons. This kind of neural synchronization may be used for efficient sensory and cognitive processing \cite{Wang1,Wang2}, and it is also correlated with pathological rhythms associated with neural diseases \cite{Disease1,Disease2,Disease3}. Here, we are interested in characterization of population synchronization of bursting neurons in terms of neural synchrony measures \cite{Order}. Bursting occurs when neuronal activity alternates, on a slow timescale, between a silent phase and an active (bursting) phase of fast repetitive spikings \cite{Rinzel1,Rinzel2,Burst1,Burst2,Burst3}. Due to the slow and fast timescales of bursting activity, bursting neurons exhibit two types of burst and spike synchronizations. Burst synchronization on the slow bursting timescale refers to a coherence between the active phase onset or offset times of bursting neurons, while spike synchronization on the fast spike timescale characterizes a coherence between intraburst spikes fired by bursting neurons  \cite{Burstsync1,Burstsync2}. Many recent studies on the burst and spike synchronizations have been made in several aspects (e.g., chaotic phase synchronization, transitions between different states of burst synchronization, effect of network topology, effect on information transmission, suppression of bursting synchronization, and effect of noise and coupling on the burst and spike synchronization) \cite{BSsync1,BSsync2,BSsync3,BSsync4,BSsync5,BSsync6,BSsync7,BSsync8,BSsync9,BSsync10,BSsync11,BSsync12,BSsync13,BSsync14,BSsync15}.

In this paper, we are interested in practical characterization of the burst and spike synchronization transitions of bursting neurons in the frequency domain. Population synchronization may be well visualized in the raster plot of neural spikes which can be obtained in experiments. Instantaneous population firing rate (IPFR), $R(t)$, which is directly obtained from the raster plot of spikes, is a realistic population quantity describing collective behaviors in both the computational and the experimental neuroscience \cite{Wang1,Sparse1,Sparse2,Sparse3,Sparse4,Sparse5,Sparse6}. This experimentally-obtainable $R(t)$ is in contrast to the ensemble-averaged potential $X_G$ which is often used as a population quantity in the computational neuroscience, because to directly get $X_G$ in real experiments is very difficult. To overcome this difficulty, instead of $X_G$, we used $R(t)$ as a population quantity, and developed a realistic order parameter, based on $R(t)$, to make practical characterization of synchronization of spiking neurons in both the computational and the experimental neuroscience \cite{Kim}. The mean square deviation of $R(t)$ plays the role of the realistic order parameter $\cal {O}$ used to determine the threshold value for the synchronization transition of spiking neurons. In this way, synchronization transition of spiking neurons may be well characterized in terms of the realistic order parameter $\cal {O}$, based on the IPFR $R(t)$.

In the field of neuroscience, power-spectral analysis of a time-series $x(t)$ (e.g., neuronal membrane potential) is often made to examine how the variance of the data $x(t)$ is distributed over the frequency components into which $x(t)$ may be decomposed \cite{Sparse1,Sparse2,Sparse3,Sparse4,Sparse5,Sparse6}. Following this conventional direction, we investigate the synchronization transitions of bursting neurons in the frequency domain. The main purpose of our works is to characterize the synchronization transitions of bursting neurons in terms of ``frequency-domain'' order parameters by extending the realistic ``time-domain'' order parameter of spiking neurons \cite{Kim} to the case of bursting neurons. This extension work on the frequency-domain order parameters is in contrast to another extension work where the synchronization transitions of bursting neurons are characterized in terms of the time-domain order parameters \cite{Kim1}. The IPFR $R(t)$ shows the whole combined population behaviors with both the slow and fast timescales. To clearly investigate the synchronization transitions of bursting neurons, we separate the slow and fast timescales of the bursting activity via frequency filtering, and decompose the IPFR $R(t)$ into $R_b(t)$ (the instantaneous population burst rate (IPBR) describing the bursting behavior) and $R_s(t)$ (the instantaneous population spike rate (IPSR) describing the intraburst spiking behavior). In presence of the burst and spike synchronizations, $R_b$ and $R_s$ exhibit regular oscillations, independently of $N$ (the number of the bursting neurons). On the other hand, in absence of the burst and spike synchronizations, $R_b$ and $R_s$ become stationary as $N$ goes to the infinity. The synchronous oscillations of $R_b$ and $R_s$ in the time domain may be well characterized by the bursting and spiking peaks of their power spectral densities. As in the case of the coherence resonance \cite{CR1}, each spectral ``resonance'' (i.e., peak) in the power spectral density may be well analyzed in terms of a ``coherence factor'' $\beta$ (i.e., a measure of spectral coherence) which is defined by a ``signal to noise'' ratio of the spectral peak height and its relative width \cite{CR2,CR3}. We also note that the signal to noise ratio has a long history of being used in neuroscience as a measure of the fidelity of signal transmission and detection by neurons and synapses \cite{CR4}. Then, the bursting and spiking coherence factors $\beta_b$ and $\beta_s$ of the bursting and spiking peaks in the power spectral densities of $R_b(t)$ and $R_s(t)$ are shown to play the role of the bursting and spiking order parameters in the frequency domain which are used to determine the bursting and spiking thresholds for the bursting and spiking transitions, respectively. We also consider another raster plot of bursting onset or offset times which visualizes the bursting behaviors more directly. From this type of raster plot, we may directly obtain the IPBR, $R_b^{(on)}(t)$ or $R_b^{(off)}(t)$, without frequency filtering. Then, the bursting onset and offset coherence factors, $\beta_b^{(on)}$ and $\beta_b^{(off)}$, of the bursting onset and offset peaks in the power spectral densities of $R_b^{(on)}(t)$ and $R_b^{(off)}(t)$ are also shown to play the role of the frequency-domain bursting order parameters for the bursting transition. The frequency-domain order parameters $ \beta_b^{(on)}$ and $\beta_b^{(off)}$ yield the same bursting threshold which is obtained through calculation of $\beta_b$, and they are more direct ones than $\beta_b$ because they may be directly obtained without frequency filtering. Consequently, all the frequency-domain bursting and spiking order parameters may be usefully used for characterization of the burst and and spike synchronization transitions of the bursting neurons in the frequency domain.

This paper is organized as follows. In Sec.~\ref{sec:HR}, as an example for characterization we describe an inhibitory network of bursting Hindmarsh-Rose (HR) neurons \cite{HR1,HR2,HR3,HR6}. In Sec.~\ref{sec:OR}, we separate the slow bursting and the fast spiking timescales via frequency filtering, and develop realistic frequency-domain bursting and spiking order parameters (i.e., the bursting and spiking coherence factors), based on the power spectral densities of the IPBR and the IPSR, which are applicable in both the computational and the experimental neuroscience.  Their usefulness for characterization of the burst and spike synchronization transitions is shown in explicit examples of bursting HR neurons. Finally, a summary is given in Section \ref{sec:SUM}.

\section{A Network of Inhibitory Bursting Hindmarsh-Rose Neurons}
\label{sec:HR}
As an example for characterization, we consider an inhibitory network of $N$ globally-coupled bursting HR neurons. The representative bursting HR neuron model was originally introduced to describe the time evolution of the membrane potential for the pond snails \cite{HR1,HR2,HR3,HR6}. The population dynamics in this inhibitory network is governed by the following set of ordinary differential equations:
\begin{eqnarray}
\frac{dx_i}{dt} &=& y_i - a x^{3}_{i} + b x^{2}_{i} - z_i +I_{DC} +D \xi_{i} -I_{syn,i}, \label{eq:CHRA} \\
\frac{dy_i}{dt} &=& c - d x^{2}_{i} - y_i, \label{eq:CHRB} \\
\frac{dz_i}{dt} &=& r \left[ s (x_i - x_o) - z_i \right], \label{eq:CHRC} \\
\frac{dg_i}{dt}&=& \alpha g_{\infty}(x_i) (1-g_i) - \beta g_i, \;\;\; i=1, \cdots, N, \label{eq:CHRD}
\end{eqnarray}
where
\begin{eqnarray}
I_{syn,i} &=& \frac{J}{N-1} \sum_{j(\ne i)}^N g_j(t) (x_i - X_{syn}), \label{eq:CHRE} \\
g_{\infty} (x_i) &=& 1/[1+e^{-(x_i-x_s^*)\delta}]. \label{eq:CHRF}
\end{eqnarray}
Here, the state of the $i$th HR neuron at a time $t$ (measured in units of milliseconds) is described by four state variables: the fast membrane potential $x_i$, the fast recovery current $y_i,$ the slow adaptation current $z_i$, and the synaptic gate variable $g_i$ representing the fraction of open synaptic ion channels. The parameters in the single HR neuron are taken as $a=1.0,$ $b=3.0,$ $c=1.0,$ $d=5.0,$ $r=0.001,$ $s=4.0,$  and $x_o=-1.6$ \cite{CR3}.

\begin{figure}
\includegraphics[width=0.6\columnwidth]{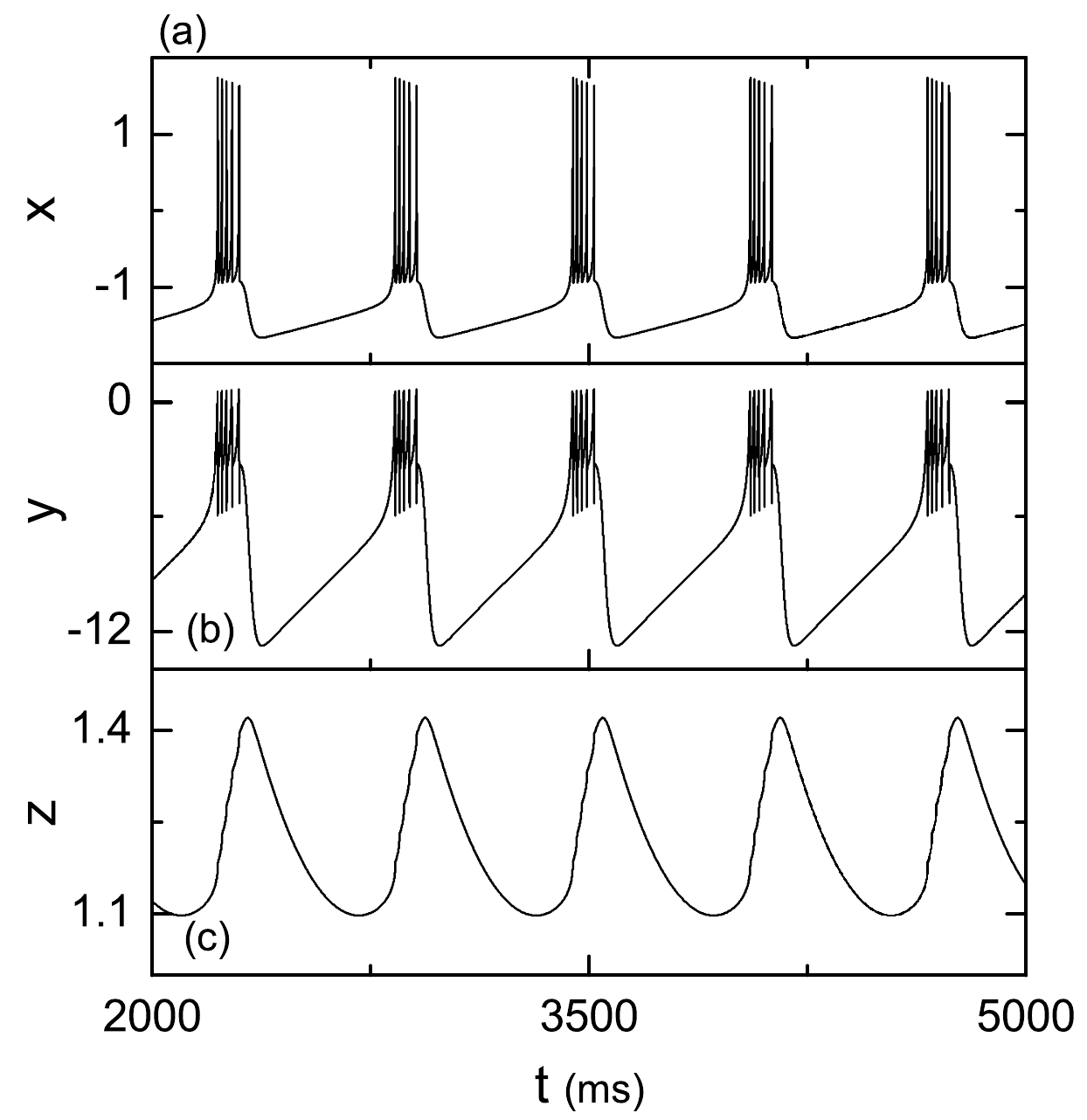}
\caption{Single HR neuron for $I_{DC}=1.3$ and $D=0$. Plots of time series of (a) the fast membrane potential $x(t)$, (b) the fast
recovery current $y(t)$, and (c) the slow adaptation current $z(t)$.
}
\label{fig:Single}
\end{figure}

Each bursting HR neuron is stimulated by using the common DC current $I_{DC}$ and an independent Gaussian white noise $\xi_i$ [see the 5th and the 6th terms in Eq.~(\ref{eq:CHRA})] satisfying $\langle \xi_i(t) \rangle =0$ and $\langle \xi_i(t)~\xi_j(t') \rangle = \delta_{ij}~\delta(t-t')$, where
$\langle\cdots\rangle$ denotes the ensemble average. The noise $\xi$ is a parametric one that randomly perturbs the strength of the applied current
$I_{DC}$, and its intensity is controlled by using the parameter $D$. As $I_{DC}$ passes a threshold $I_{DC}^* (\simeq 1.26)$ in the absence of noise, each single HR neuron exhibits a transition from a resting state to a bursting state. Throughout this paper, we consider the suprathreshold case of $I_{DC}=1.3$
where each HR neuron exhibits spontaneous bursting activity without noise. Figures \ref{fig:Single}(a)-\ref{fig:Single}(b) show the time series of the fast membrane potential $x(t)$ and the fast recovery current $y(t)$, while the time series of the slow adaptation current $z(t)$ is shown in Fig.~\ref{fig:Single}(c). As seen well in the time series of $x$ and $y$, the bursting activity alternates, on a slow timescale, between a silent phase and an active (bursting) phase of fast repetitive spikings. For this case, the slow bursting timescale is $\tau_b \simeq 609$ ms [corresponding to the slow bursting frequency $f_b$ $(\simeq 1.6$ Hz)], while the fast spiking timescale is $\tau_s \simeq 18.2$ ms [corresponding to the fast spiking frequency $f_s$ $(\simeq 55$ Hz)].

The last term in Eq.~(\ref{eq:CHRA}) represents the synaptic coupling of the network. $I_{syn,i}$ of Eq.~(\ref{eq:CHRE}) represents a synaptic current injected into the $i$th neuron. Here the coupling strength is controlled by the parameter $J$ and $X_{syn}$ is the synaptic reversal potential. Here, we use $X_{syn}=-2$ for the inhibitory synapse. The synaptic gate variable $g$ obeys the 1st order kinetics of Eq.~(\ref{eq:CHRD}) \cite{KI1,KI2}. Here, the normalized concentration of synaptic transmitters, activating the synapse, is assumed to be an instantaneous sigmoidal function of the membrane potential with a threshold $x_s^*$ in Eq.~(\ref{eq:CHRF}), where we set $x_s^*=0$ and $\delta=30$ \cite{HR5}.
The transmitter release occurs only when the neuron emits a spike (i.e., its potential $x$ is larger than $x^*_s$). For the inhibitory GABAergic synapse (involving the $\rm{GABA_A}$ receptors), the synaptic channel opening rate, corresponding to the inverse of the synaptic rise time $\tau_r$, is $\alpha=10$ ${\rm ms}^{-1}$, and the synaptic closing rate $\beta$, which is the inverse of the synaptic decay time $\tau_d$, is $\beta=0.1$ ${\rm ms}^{-1}$ \citep{GABA1,GABA2}. Hence, $I_{syn}$ rises fast and decays slowly.

Numerical integration of Eqs.~(\ref{eq:CHRA})-(\ref{eq:CHRD}) is done using the Heun method \cite{SDE} (with the time step $\Delta t=0.01$ ms).
For each realization of the stochastic process, we choose a random initial point $[x_i(0),y_i(0),z_i(0),g_i(0)]$ for the $i$th $(i=1,\dots, N)$ neuron with uniform probability in the range of $x_i(0) \in (-2,2)$, $y_i(0) \in (-16,0)$, $z_i(0) \in (1.1,1.4)$, and $g_i(0) \in (0,1)$.

\section{Frequency-domain order parameters for the burst and spike synchronization transitions}
\label{sec:OR}
In this section, we extend the realistic order parameter of spiking neurons to the case of bursting neurons for characterization of population synchronization transition in the frequency domain. For our aim, we separate the slow bursting and the fast spiking timescales through frequency filtering, and decompose the IPFR $R(t)$ into the IPBR $R_b(t)$ (describing  the bursting behavior) and the IPSR $R_s(t)$ (describing the intraburst spiking behavior). Then, we develop realistic frequency-domain bursting and spiking order parameters, based on the power spectral densities of the IPBR $R_b(t)$ and the IPSR $R_s(t)$, and show their usefulness for characterization of the burst and spike synchronization transitions in explicit examples of bursting HR neurons.

\begin{figure}
\includegraphics[width=0.8\columnwidth]{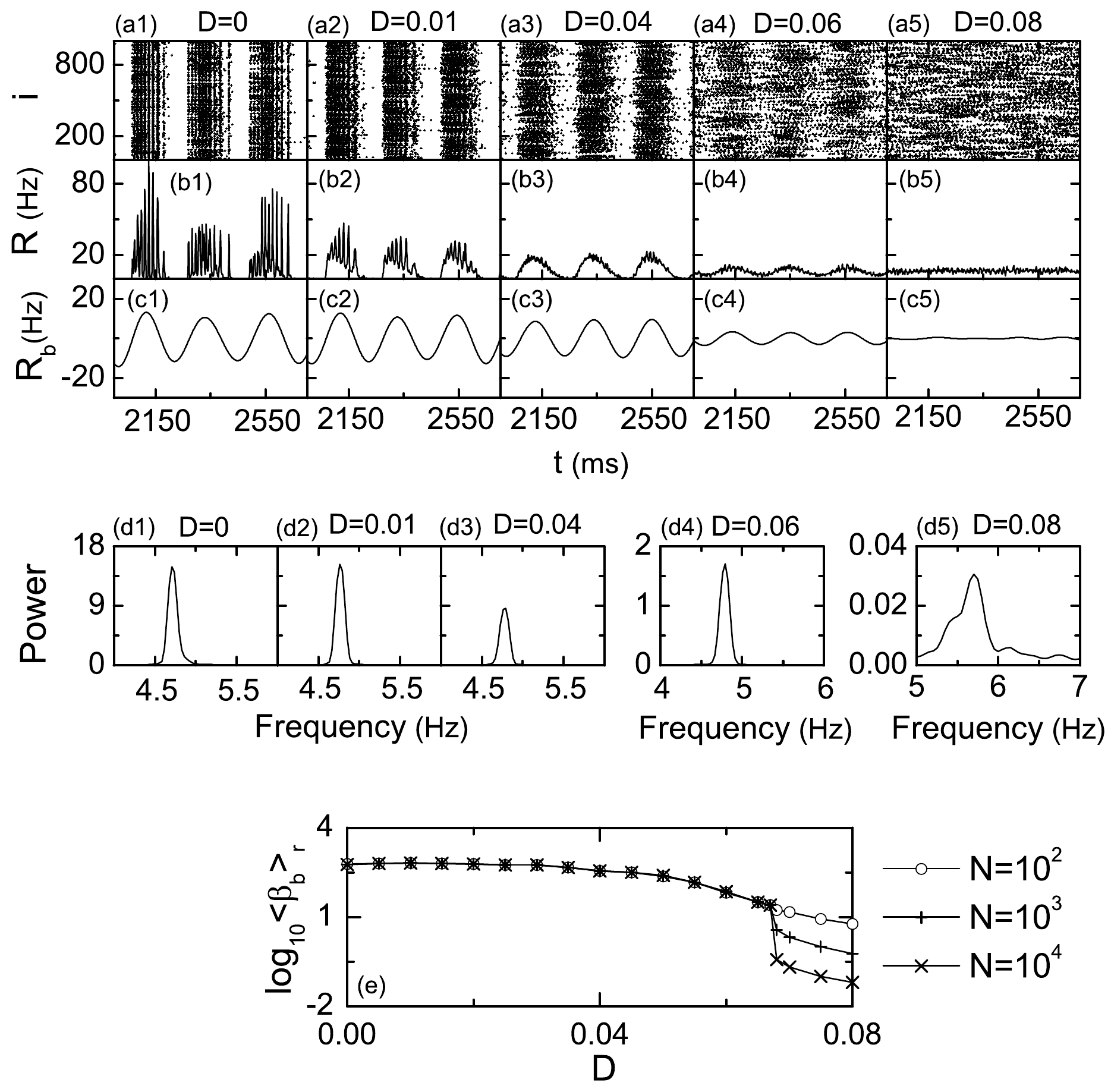}
\caption{Population bursting states for various values of $D$ and determination of the bursting noise threshold $D^*_b$ in an inhibitory ensemble of $N$ globally-coupled bursting HR neurons for $I_{DC}=1.3$ and $J=0.3$: synchronized bursting states for $D=0,$ 0.01, 0.04, and 0.06, and unsynchronized bursting state for $D=0.08$. $N=10^3$ except for the case of (e). (a1)-(a5) Raster plots of neural spikes, (b1)-(b5) time series of IPFR kernel estimate $R(t)$ (the band width $h$ of the Gaussian kernel function is 1 ms), (c1)-(c5) time series of band-pass filtered IPBR $R_b(t)$ [lower and higher cut-off frequencies of 3 Hz (high-pass filter) and 7 Hz (low-pass filter)], and (d1)-(d5) one-sided power spectra of $\Delta R_b(t)$ $[=R_b(t)- \overline{R_b(t)}]$ with mean-squared amplitude normalization. Each power spectrum in (d1)-(d5) is made of $2^{15}$ data points and it is smoothed by the Daniell filters of lengths 3 and 5. (e) Plots of realistic frequency-domain bursting order parameter ${\langle \beta_b \rangle}_r$ versus $D$: ${\langle \beta_b \rangle}_r$ is obtained through average over 20 realizations for each $D$.
}
\label{fig:Bursting1}
\end{figure}

\begin{figure}
\includegraphics[width=0.8\columnwidth]{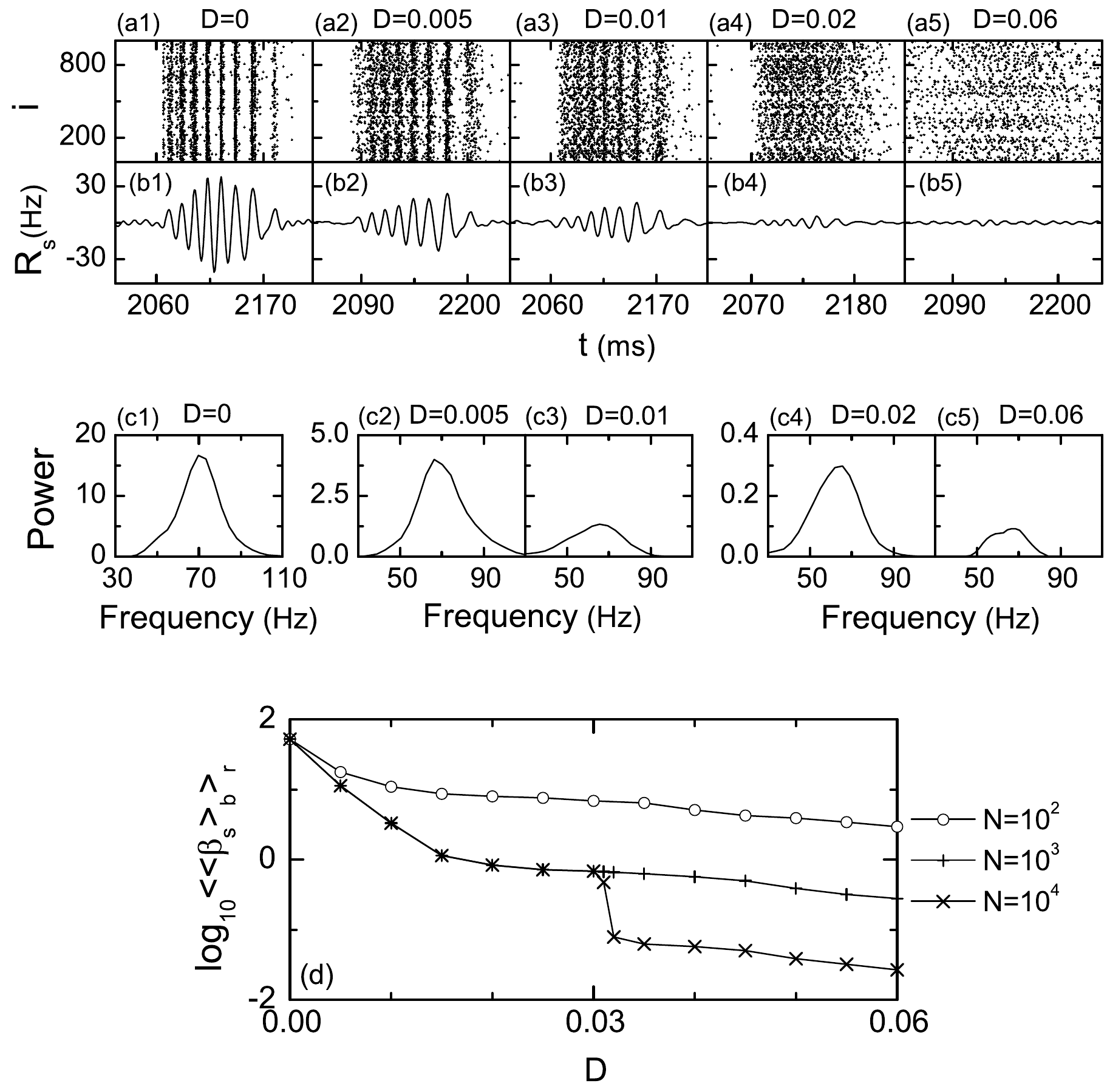}
\caption{Population intraburst spiking states for various values of $D$ and determination of the bursting noise threshold $D^*_s$ in an inhibitory ensemble of $N$ globally-coupled bursting HR neurons for $I_{DC}=1.3$ and $J=0.3$: synchronized spiking states for $D=0,$ 0.005, 0.01, and 0.02, and unsynchronized spiking state for $D=0.06$. $N=10^3$ except for the case of (d). (a10-(a5) Raster plots of neural spikes and (b1)-(b5) time series of the band-pass filtered IPSR $R(t)$ [lower and higher cut-off frequencies of 30 Hz (high-pass filter) and 90 Hz (low-pass filter)] in the 1st global bursting cycle of the IPBR $R_b(t)$ (after the transient time of $2 \times 10^3$ ms) for each $D$. (c1)-(c5) One-sided power spectra of $\Delta R_s(t)$ $[=R_s(t)- \overline{R_s(t)}]$ with mean-squared amplitude normalization. Each power spectrum in (c1)-(c5) is made of $2^{8}$ data points for each global bursting cycle of $R_b(t)$ and it is smoothed by the Daniell filters of lengths 3 and 5. (d) Plots of realistic frequency-domain spiking order parameter ${\langle {\langle \beta_s \rangle}_b \rangle}_r$ versus $D$; ${\langle {\langle \beta_s \rangle}_b \rangle}_r$ is obtained through double-averaging over the 20 bursting cycles and the 20 realizations.
}
\label{fig:Spiking}
\end{figure}

As an example for characterization, we consider an inhibitory network of $N$ globally-coupled bursting HR neurons, and characterize the synchronization transitions of bursting HR neurons
in the frequency domain by varying the noise intensity $D$. To compare our results in the frequency domain with those in the time domain, we fix the DC current strength $I_{DC}$ and the coupling strength $J$ at $I_{DC}=1.3$ and $J=0.3$, as in the time-domain work \cite{Kim1}. In computational neuroscience, a population-averaged global potential,
\begin{equation}
 X_G (t) = \frac {1} {N} \sum_{i=1}^{N} x_i(t),
\label{eq:GPOT}
\end{equation}
is often used for describing emergence of population synchronization. In this study, we consider the population behaviors after the transient time of $2 \times 10^3$ ms. Although the global potential $X_G$ is an important ensemble-averaged quantity to describe synchronization in computational neuroscience, it is practically difficult to directly get $X_G$ in real experiments. To overcome this difficulty, instead of $X_G$, we use the IPFR which is an experimentally-obtainable population quantity used in both the experimental and the computational neuroscience \cite{Wang1,Sparse1,Sparse2,Sparse3,Sparse4,Sparse5,Sparse6}. The IPFR is obtained from the raster plot of spikes which is a collection of spike trains of individual neurons. Such raster plots of spikes, where population synchronization may be well visualized, are fundamental data in the experimental neuroscience. The raster plots of spikes in Figs.~\ref{fig:Bursting1}(a1)-\ref{fig:Bursting1}(a5) show population states for various values of noise intensity $D$. To get a smooth IPFR from the raster plot of spikes, we employ the kernel density estimation (kernel smoother) \cite{Kernel}. Each spike in the raster plot is convoluted (or blurred) with a kernel function $K_h(t)$ to get a smooth estimate of IPFR, $R(t)$:
\begin{equation}
R(t) = \frac{1}{N} \sum_{i=1}^{N} \sum_{s=1}^{n_i} K_h (t-t_{s}^{(i)}),
\label{eq:IPSRK}
\end{equation}
where $t_{s}^{(i)}$ is the $s$th spiking time of the $i$th neuron, $n_i$ is the total number of spikes for the $i$th neuron, and we use a Gaussian
kernel function of band width $h$:
\begin{equation}
K_h (t) = \frac{1}{\sqrt{2\pi}h} e^{-t^2 / 2h^2}, ~~~~ -\infty < t < \infty.
\label{eq:Gaussian}
\end{equation}
Figures \ref{fig:Bursting1}(b1)-\ref{fig:Bursting1}(b5) show smooth IPFR kernel estimates $R(t)$ of band width $h=1$ ms for $D=0$, 0.01, 0.04, 0.06 and 0.08, respectively. For $D=0$, clear ``bursting bands,'' each of which is composed of ``stripes'' of spikes, appear successively at nearly regular time intervals [see Fig.~\ref{fig:Bursting1}(a1)]; a magnified 1st intraburst band is given in Fig.~\ref{fig:Spiking}(a1). For the case of $D=0$, both the burst synchronization [synchrony on the slow bursting timescale $\tau_b$ ($\simeq 215$ ms)] and the spike synchronization [synchrony on the fast spike timescale $\tau_s$ $(\simeq 14.6$ ms)] occur in each bursting band. As a result of this complete synchronization, the IPFR kernel estimate $R(t)$ shows a bursting activity [i.e., fast spikings appear on a slow wave in $R(t)$], as shown in Fig.~\ref{fig:Bursting1}(b1). However, as $D$ is increased, loss of spike synchronization occurs in each bursting band because spiking stripes become smeared due to a destructive role of noise. As an example, see the case of $D=0.01$ where the raster plot of spikes and the IPFR kernel estimate $R(t)$ are shown in Figs.~\ref{fig:Bursting1}(a2) and \ref{fig:Bursting1}(b2), respectively. The magnified 1st bursting band in Fig.~\ref{fig:Spiking}(a3) shows smearing of the spiking stripes well. Consequently, the amplitude of $R(t)$ decreases, as shown in Fig.~\ref{fig:Bursting1}(b2). As $D$ is further increased and passes a spiking noise threshold $D^*_s$ $(\simeq 0.032)$, complete loss of spike synchronization occurs in each bursting band. Then, only the burst synchronization (without spike synchronization) occurs, as shown in the case of $D=0.04$ in Figs.~\ref{fig:Bursting1}(a3) and \ref{fig:Bursting1}(b3). For this case, $R(t)$ shows a slow-wave oscillation without spikes. With increase in $D$, such ``incoherent'' bursting bands become more and more smeared, and hence the degree of burst synchronization decreases [e.g., see the case of $D=0.06$ in Fig.~\ref{fig:Bursting1}(a4)]. As a result, the amplitude of $R(t)$ is further decreased, as shown in Fig.~\ref{fig:Bursting1}(b4) for $D=0.06$. With further increasing $D$, incoherent bursting bands begin to overlap, which eventually results in the complete loss of burst synchronization as $D$ passes another larger bursting noise threshold $D^*_b$ $(\simeq 0.068)$. Consequently, for $D> D^*_b$, completely unsynchronized states with nearly stationary $R(t)$ appear, as shown in the case of $D=0.08$ in Figs.~\ref{fig:Bursting1}(a5) and \ref{fig:Bursting1}(b5).

The (above) IPFR kernel estimate $R(t)$ is a population quantity describing the ``whole'' combined collective behaviors of bursting neurons with both the slow bursting and the fast spiking timescales. Through frequency filtering, we separate the slow and the fast timescales, and decompose the IPFR kernel estimate $R(t)$ into the IPBR $R_b(t)$ and the IPSR $R_s(t)$ for more clear investigation of the burst and spike synchronizations. Through band-pass filtering of $R(t)$ [with the lower and the higher cut-off frequencies of 3 Hz (high-pass filter) and 7 Hz (low-pass filer)], we get the regularly-oscillating IPBR $R_b(t)$ (containing only the slow wave without spikes) in Figs.~\ref{fig:Bursting1}(c1)-\ref{fig:Bursting1}(c5) for $D=0$, 0.01, 0.04, 0.06, and 0.08. As $D$ is increased, the amplitude of $R_b(t)$ decreases gradually, and eventually $R_b(t)$ becomes nearly stationary when $D$ passes the bursting noise threshold $D^*_b$ $(\simeq 0.068)$. We note that synchronous oscillations of $R_b(t)$ in the time domain are characterized by the bursting peaks in the power spectral densities of $\Delta R_b(t)$ $[=R_b(t)- \overline{R_b(t)}]$, where the overline represents the time average. Figures \ref{fig:Bursting1}(d1)-\ref{fig:Bursting1}(d5) show distinct bursting peaks in the power spectra of $\Delta R_b(t)$; each power spectrum is made of $2^{15}$ data points and smoothed through the Daniell filters of length 3 and 5 \cite{PS}. Then, each bursting peak may be analyzed well in terms of a bursting coherence factor $\beta_b$ defined by the product of the height $H_p$ and the $Q$ factor of the peak \cite{CR1,CR2,CR3}:
\begin{equation}
\beta_b = H_p~Q; Q = f_p / \Delta f_p.
\end{equation}
Here, $f_p$ and $\Delta f_p$ are the frequency of the bursting peak and the width of the bursting peak at the height of $e^{-1/2}~h$, respectively. For more accurate results, we repeat the process to get the bursting coherence factor $\beta_b$ for multiple realizations. Thus, we obtain ${\langle \beta_b \rangle}_r$ (average bursting coherence factor) through average over 20 realizations. Figure \ref{fig:Bursting1}(e) shows plots of the average bursting coherence factor ${\langle \beta_b \rangle}_r$ versus $D$. For $D < D^*_b$ $(\simeq 0.068$), synchronized bursting states exist because the values of ${\langle \beta_b \rangle}_r$ become saturated to non-zero limit values in the thermodynamic limit of $N \rightarrow \infty$ (i.e., bursting peaks persist, independently of $N$). However, as $D$ passes the bursting noise threshold $D^*_b$, the average bursting coherence factor ${\langle \beta_b \rangle}_r$ tends to zero as $N \rightarrow \infty$ (i.e., eventually bursting peaks disappear in the thermodynamic limit), and hence a transition to unsynchronized bursting states occurs because the noise spoils the burst synchronization completely. In this way, the average bursting coherence factor ${\langle \beta_b \rangle}_r$ describes the burst synchronization transition well in the frequency domain, and hence it plays the role of the realistic frequency-domain bursting order parameter for the bursting transition (i.e., one can determine the bursting noise threshold $D^*_b$ through calculation of ${\langle \beta_b \rangle}_r$). This frequency-domain bursting order parameter ${\langle \beta_b \rangle}_r$ is in contrast to the time-domain bursting order parameter, based on the time-averaged fluctuation of the IPBR $R_b(t)$ \cite{Kim1}. In spite of their difference, calculations of both the frequency-domain and the time-domain bursting order parameters result in the same bursting noise threshold $D^*_b$ (compare Fig.~\ref{fig:Bursting1}(e) with Fig.~3(a) in \cite{Kim1}). Consequently, the frequency-domain bursting order parameter may be used effectively to determine $D^*_b$ for the bursting transition, like the case of the time-domain bursting order parameter.

From now on, we investigate the intraburst spike synchronization transition of bursting HR neurons in the frequency domain by varying the noise intensity $D$.
Figures \ref{fig:Spiking}(a1)-\ref{fig:Spiking}(a5) and Figures \ref{fig:Spiking}(b1)-\ref{fig:Spiking}(b5) show the raster plots of intraburst spikes and the corresponding (band-pass filtered) IPSR $R_s(t)$ during the 1st global bursting cycle of the IPBR $R_b(t)$, respectively for various values of $D$: synchronized spiking states for $D=0$, 0.005, 0.01, and 0.02, and unsynchronized spiking state for $D=0.06$. Here, the IPSRs $R_s(t)$ are obtained through band-pass filtering of the IPFR kernel estimate $R(t)$ [with the lower and the higher cut-off frequencies of 30 Hz (high-pass filter) and 90 Hz (low-pass filer)]. Then, the intraburst spike synchronization may be well described in terms of $R_s(t)$. For $D=0$, clear 8 spiking stripes (composed of spikes and indicating population spike synchronization) appear in the intraburst band of the 1st global bursting cycle of $R_b(t)$ in Fig.~\ref{fig:Spiking}(a1), and the band-pass filtered IPSR $R_s(t)$ shows only the fast spiking oscillations (without a slow wave) with the population spiking frequency $f_s$ $(\simeq 68.5$ Hz) in Fig.~\ref{fig:Spiking}(b1). However, as $D$ is increased, spiking stripes in the intraburst band become more and more smeared (e.g., see the cases of $D=0.005$, 0.01, and 0.02). Consequently, the amplitude of $R_s(t)$ decreases due to loss of spike synchronization. Eventually, when $D$ passes the spiking noise threshold $D^*_s$ $(\simeq 0.032)$, spikes become completely scattered within the intraburst band, and $R_s(t)$ becomes nearly stationary. Consequently, for $D> D^*_s$, complete loss of spike synchronization occurs in the intraburst band, as shown in Fig.~\ref{fig:Spiking}(b5) for $D=0.06$.
Figures \ref{fig:Spiking}(c1)-\ref{fig:Spiking}(c5) show the power spectra of $\Delta R_s(t)$ $[=R_s(t)- \overline{R_s(t)}]$ in the 1st global bursting cycle of $R_b(t)$: each power spectrum is made of $2^{8}$ data points and smoothed through the Daniell filters of length 3 and 5. Spiking peaks in their power spectra are analyzed in terms of the spiking coherence factors $\beta_s$ (defined by the product of the height $H_p$ and the $Q$ factor of the peak). For more accurate results, we repeat the process to get $\beta_s$ for multiple realizations. In each realization we follow the 20 global bursting cycles of $R_b(t)$, and get the double-averaged spiking coherence factor ${\langle {\langle \beta_s \rangle}_b \rangle}_r$ through average over 20 realizations. Figure \ref{fig:Spiking}(d) shows plots of the double-averaged spiking coherence factor ${\langle {\langle \beta_s \rangle}_b \rangle}_r$ versus $D$. For $D < D^*_s$ $(\simeq 0.032$), synchronized spiking states exist because the values of ${\langle {\langle \beta_s \rangle}_b \rangle}_r$ become saturated to non-zero limit values as $N \rightarrow \infty$ (i.e., spiking peaks persist, irrespectively of $N$). However, when $D$ passes the spiking noise threshold $D^*_s$, ${\langle {\langle \beta_s \rangle}_b \rangle}_r$ tends to zero in the thermodynamic limit of $N \rightarrow \infty$ (i.e., eventually spiking peaks disappear in the thermodynamic limit), and hence a transition to unsynchronized spiking states occurs because the noise spoils the intraburst spike synchronization completely. In this way, the double-averaged spiking coherence factor ${\langle {\langle \beta_s \rangle}_b \rangle}_r$ describes the intraburst spike synchronization transition well in the frequency domain, and hence it plays the role of the realistic frequency-domain spiking order parameter for the spiking transition (i.e., one can determine the spiking noise threshold $D^*_s$ through calculation of ${\langle {\langle \beta_s \rangle}_b \rangle}_r$). This frequency-domain spiking order parameter is also in contrast to the time-domain spiking order parameter, based on the time-averaged fluctuation of the IPSR $R_s(t)$ \cite{Kim1}. We also note that both the frequency-domain and  the time-domain spiking order parameters yield the same spiking noise threshold $D^*_s$ (compare Fig.~\ref{fig:Spiking}(d) with Fig.~6(d) in \cite{Kim1}). Consequently, the frequency-domain spiking order parameter may also be used effectively to determine $D^*_s$ for the spiking transition, as in the case of the time-domain spiking order parameter.

\begin{figure}
\includegraphics[width=0.6\columnwidth]{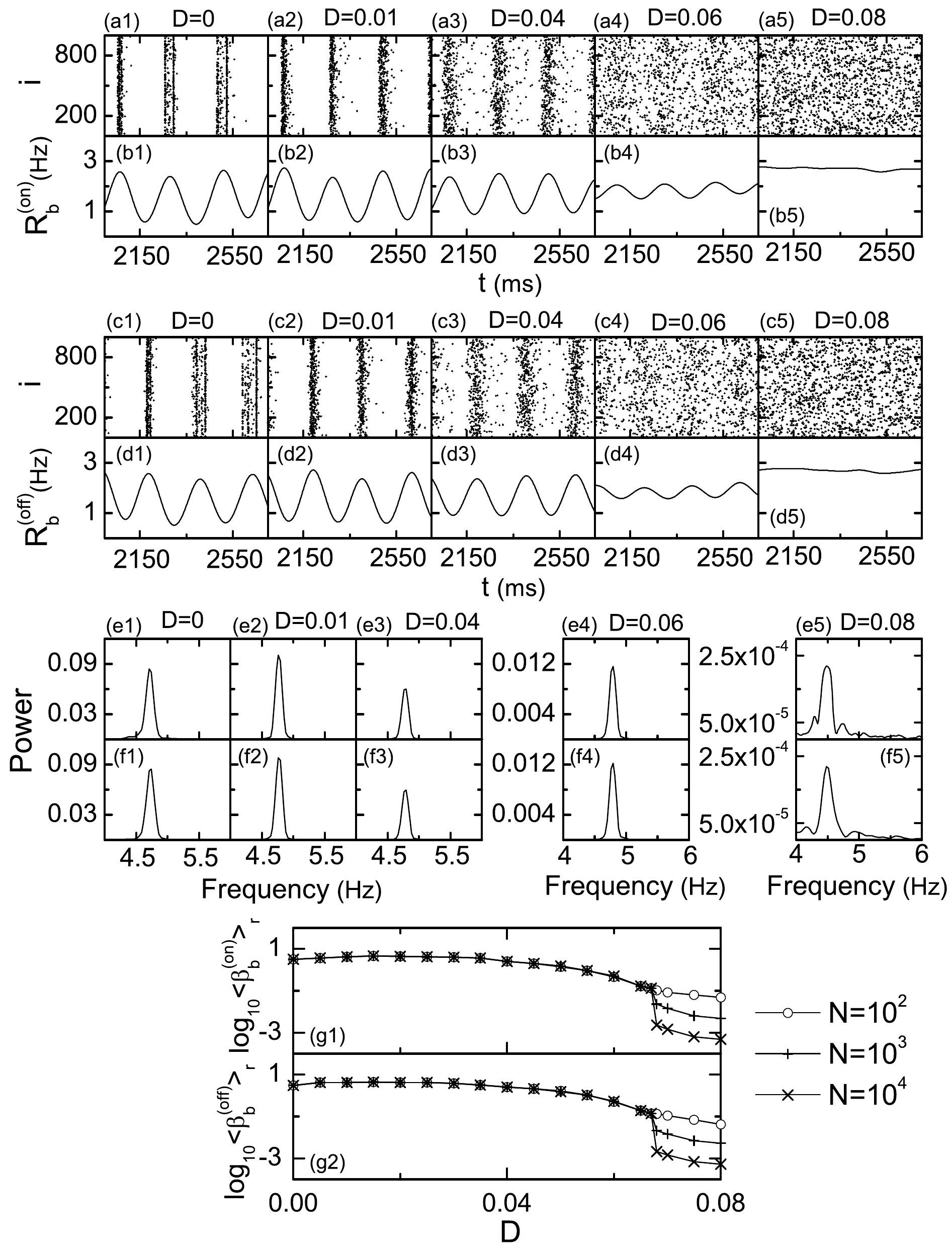}
\caption{Population bursting states represented by the bursting onset and offset times for various values of $D$ and determination of the bursting noise threshold $D^*_b$ in an inhibitory ensemble of $N$ globally-coupled bursting HR neurons for $I_{DC}=1.3$ and $J=0.3$: synchronized bursting states for $D=0,$ 0.01, 0.04, and 0.06, and unsynchronized bursting state for $D=0.08$. $N=10^3$ except for the cases of (g1) and (g2). (a1)-(a5) Raster plots of the bursting onset times and (b1)-(b5) time series of the IPBR $R_b^{(on)}(t)$ (the band width $h$ of the Gaussian kernel function is 50 ms). (c10-(c5) Raster plot of the bursting offset times and (d1)-(d5) time series of the IPBR $R_b^{(off)}(t)$ (the band width $h$ of the Gaussian kernel function is 50 ms). (e1)-(e5) One-sided power spectra of $\Delta R_b^{(on)}(t)$ $[=R_b^{(on)}(t)- \overline{R_b^{(on)}(t)}]$ with mean-squared amplitude normalization
and (f1)-(f5) one-sided power spectra of $\Delta R_b^{(off)}(t)$ $[=R_b^{(off)}(t)- \overline{R_b^{(off)}(t)}]$ with mean-squared amplitude normalization. Each power spectrum is made of $2^{15}$ data points and it is smoothed by the Daniell filters of lengths 3 and 5. Plots of realistic frequency-domain bursting order parameters (g1) ${\langle \beta_b^{on)} \rangle}_r$ and (g2) ${\langle \beta_b^{off)} \rangle}_r$ versus $D$: ${\langle \beta_b^{(on)} \rangle}_r$ and  ${\langle \beta_b^{(off)} \rangle}_r$ are obtained through average over 20 realizations for each $D$.
}
\label{fig:Bursting2}
\end{figure}

Finally, we consider another raster plot of bursting onset or offset times for more direct visualization of bursting behavior.
[At the onset (offset) times of the $i$th bursting HR neuron, its individual potential $x_i$ passes the threshold of $x^*_b = -1$ from below (above).] Without frequency filtering, we can directly obtain the IPBR kernel estimate, $R_b^{(on)}(t)$ [$R_b^{(off)}(t)$] from the raster plot of the bursting onset (offset) times. Figures \ref{fig:Bursting2}(a1)-\ref{fig:Bursting2}(a5) show the raster plots of the bursting onset times for various values of $D$, while the raster plots of the bursting offset times are shown in Figs.~\ref{fig:Bursting2}(c1)-\ref{fig:Bursting2}(c5). From these raster plots of the bursting onset (offset) times, we obtain smooth IPBR kernel estimates, $R_b^{(on)}(t)$ [$R_b^{(off)}(t)$], of band width $h=50$ ms in Figs.~\ref{fig:Bursting2}(b1)[(d1)]-\ref{fig:Bursting2}(b5)[(d5)] for $D=0$, 0.01, 0.04, 0.06, and 0.08. For $D=0$, clear bursting ``stripes'' [composed of bursting onset (offset) times and indicating burst synchronization] appear successively at nearly regular time intervals; the bursting onset and offset stripes are time-shifted [see Figs.~\ref{fig:Bursting2}(a1) and \ref{fig:Bursting2}(c1)]. The corresponding IPBR kernel estimates, $R_b^{(on)}(t)$ and $R_b^{(off)}(t)$, for $D=0$ show regular oscillations with the same population bursting frequency $f_b$ $(\simeq 4.7$ Hz), although they are phase-shifted [see Figs.~\ref{fig:Bursting2}(b1) and \ref{fig:Bursting2}(d1)]. With increasing $D$, the bursting onset and offset stripes in the raster plots become smeared and begin to overlap, and thus the degree of the burst synchronization decreases. As a result, the amplitudes of both $R_b^{(on)}(t)$ and $R_b^{(off)}(t)$ decrease gradually (e.g., see the cases of $D=0.01$, 0.04, and 0.06). Eventually, as $D$ passes the bursting noise threshold $D^*_b$ $(\simeq 0.068)$, bursting onset and offset times become completely scattered in the raster plots, and the corresponding IPBR kernel estimates, $R_b^{(on)}(t)$ and $R_b^{(off)}(t)$, become nearly stationary, as shown in Figs.~\ref{fig:Bursting2}(b5) and \ref{fig:Bursting2}(d5) for $D=0.08$. Figures \ref{fig:Bursting2}(e1)-\ref{fig:Bursting2}(e5) show the power spectra of $\Delta R_b^{(on)}(t)$ $[=R_b^{(on)}(t)- \overline{R_b^{(on)}(t)}]$, while Figures \ref{fig:Bursting2}(f1)-\ref{fig:Bursting2}(f5) show the power spectra of $\Delta R_b^{(off)}(t)$ $[=R_b^{(off)}(t)- \overline{R_b^{(off)}(t)}]$; each power spectrum is made of $2^{15}$ data points and smoothed through the Daniell filters of length 3 and 5. Bursting onset and offset peaks in these power spectra are analyzed in terms of the bursting onset and offset coherence factors $\beta_b^{(on)}$ and $\beta_b^{(off)}$ (each coherence factor of a peak is defined by the product of the height $h$ and the $Q$ factor of the peak). For more accurate results, we repeat the process to obtain $\beta_b^{(on)}$ and $\beta_b^{(off)}$ for multiple realizations. Thus, we obtain ${\langle \beta_b^{(on)} \rangle}_r$ (average bursting onset coherence factor) and ${\langle \beta_b^{(off)} \rangle}_r$ (average bursting offset coherence factor) through average over 20 realizations. Figures \ref{fig:Bursting2}(g1) and \ref{fig:Bursting2}(g2) show plots of the average bursting onset and offset coherence factor ${\langle \beta_b^{(on)} \rangle}_r$ and ${\langle \beta_b^{(off)} \rangle}_r$ versus $D$, respectively. As in the case of the (above) average bursting coherence factor ${\langle \beta_b \rangle}_r$, when passing the same bursting noise threshold $D^*_b$ $(\simeq 0.068)$, both the average bursting onset and offset coherence factors ${\langle \beta_b^{(on)} \rangle}_r$ and ${\langle \beta_b^{(off)} \rangle}_r$ go to zero as $N \rightarrow \infty$ (i.e., eventually bursting onset and offset peaks disappear in the thermodynamic limit), and hence a transition to burst unsynchronization occurs for $D > D^*_b$, because the noise breaks up the burst synchronization completely. In this way, both the average bursting onset and offset coherence factors, ${\langle \beta_b^{(on)} \rangle}_r$ and ${\langle \beta_b^{(off)} \rangle}_r$, describe the burst synchronization transition well in the frequency domain, and hence they also play the role of the realistic frequency-domain bursting order parameters for the bursting transition together with ${\langle \beta_b \rangle}_r$. These frequency-domain bursting order parameters are also in contrast to the time-domain bursting order parameters, based on the time-averaged fluctuations of
$R_b^{(on)}(t)$ and $R_b^{(off)}(t)$ \cite{Kim1}. We note that both the frequency-domain and the time-domain bursting order parameters yield the same bursting noise threshold $D^*_b$ (compare Figs.~\ref{fig:Bursting2}(g1) and \ref{fig:Bursting2}(g2) with Figs.~3(b) and 3(c) in \cite{Kim1}). Consequently, along with
${\langle \beta_b \rangle}_r$, the frequency-domain bursting order parameters, ${\langle \beta_b^{(on)} \rangle}_r$ and ${\langle \beta_b^{(off)} \rangle}_r$, may also be used effectively to determine $D^*_b$ for the bursting transition, as in the case of the time-domain bursting order parameters.

\section{Summary} \label{sec:SUM}
We have extended the realistic time-domain order parameter of spiking neurons to the case of bursting neurons. Their usefulness for characterization of the burst and spike synchronization transitions in the frequency domain has been shown in explicit examples of bursting HR neurons by varying the noise intensity $D$. Population synchronization may be well visualized in the raster plot of neural spikes which may be obtained in experiments. The IPFR kernel estimate $R(t)$, which is obtained from the raster plot of spikes, is a realistic collective quantity describing the whole combined population behaviors with the slow bursting and the fast spiking timescales. Through frequency filtering, we have decomposed the IPFR kernel estimate $R(t)$ into the IPBR $R_b(t)$ and the IPSR $R_s(t)$. We note that both $R_b(t)$ and $R_s(t)$ may be used to effectively characterize the burst and spike synchronizations, respectively. For synchronous cases, oscillations of $R_b$ and $R_s$ in the time domain are characterized by the bursting and spiking peaks in their power spectral densities. Similar to the case of coherence resonance, each spectral resonance (i.e., peak) may be well analyzed in terms of the coherence factor $\beta$, defined by a ``signal to noise'' ratio of the spectral peak height and its relative width. The average bursting and spiking coherence factors ${\langle \beta_b \rangle}_r$ and ${\langle {\langle \beta_s \rangle}_b \rangle}_r$ of the bursting and spiking peaks in the power spectral densities of $\Delta R_b$ and $\Delta R_s$ have been found to play the role of the frequency-domain bursting and spiking order parameters for the burst and spike synchronization transitions, respectively. Through calculation of ${\langle \beta_b \rangle}_r$ and ${\langle {\langle \beta_s \rangle}_b \rangle}_r$, we have determined the noise bursting and spiking thresholds, $D^*_b$ and $D^*_s$, beyond which the burst and spike synchronizations break up, respectively. For more direct visualization of bursting behavior, we consider another raster plot of bursting onset or offset times, from which the IPBR, $R_b^{(on)}(t)$ or $R_b^{(off)}(t)$, can be directly obtained without frequency filtering. Then, the average bursting onset and offset coherence factors, ${\langle \beta_b^{(on)} \rangle}_r$ and ${\langle \beta_b^{(off)} \rangle}_r$ of the bursting onset and offset peaks in the power spectral densities of $\Delta R_b^{(on)}(t)$ and $\Delta R_b^{(off)}(t)$ have also been shown to play the role of the frequency-domain bursting order parameters for the bursting transition. These frequency-domain order parameters ${\langle \beta_b^{(on)} \rangle}_r$ and ${\langle \beta_b^{(off)} \rangle}_r$ yield the same bursting noise threshold $D^*_b$ which is obtained via calculation of ${\langle \beta_b \rangle}_r$, and they are more direct ones than ${\langle \beta_b \rangle}_r$ because they may be directly obtained without frequency filtering. We also note that all these bursting and spiking noise thresholds are the same as those obtained through calculations of the time-domain bursting and spiking order parameters \cite{Kim1}. Consequently, the frequency-domain bursting and spiking order parameters may be usefully used for characterizing the burst and spike synchronization transitions of the bursting neurons, as in the case of the time-domain bursting and spiking order parameters.

\begin{acknowledgments}
This research was supported by Basic Science Research Program through the National Research Foundation of Korea (NRF) funded by the Ministry of Education (Grant No. 2013057789).
\end{acknowledgments}

\newpage

\newpage

\newpage

\newpage

\end{document}